\documentclass[showpacs,aps,prb,superscriptaddress,twocolumn,10pt,amsmath]{revtex4-1}
\usepackage{graphicx}
\newcommand{\abs}[1]{\left| #1 \right|} 
\newcommand{\ket}[1]{\left| #1 \right>}
\newcommand{\pd}[2]{\frac{\partial #1}{\partial #2}} 

\begin{document}

\title{Magnetic flux tuning of Fano-Kondo interplay in a parallel double quantum dot system}
\author{R. R. Agundez}
\affiliation{Kavli Institute of Nanoscience, Delft University of Technology, Lorentzweg 1, 2628 CJ Delft, The Netherlands}
\author{J. Verduijn}
\affiliation{Kavli Institute of Nanoscience, Delft University of Technology, Lorentzweg 1, 2628 CJ Delft, The Netherlands}
\affiliation{Centre for Quantum Computation and Communication Technology, University of New South Wales, Sydney NSW 2052, Australia}
\author{S. Rogge}
\affiliation{Kavli Institute of Nanoscience, Delft University of Technology, Lorentzweg 1, 2628 CJ Delft, The Netherlands}
\affiliation{Centre for Quantum Computation and Communication Technology, University of New South Wales, Sydney NSW 2052, Australia}
\author{M. Blaauboer}
\affiliation{Kavli Institute of Nanoscience, Delft University of Technology, Lorentzweg 1, 2628 CJ Delft, The Netherlands}

\begin{abstract}
We investigate the Fano-Kondo interplay in an Aharonov-Bohm ring with an embedded non-interacting quantum dot and a Coulomb interacting quantum dot. Using a slave-boson mean-field approximation we diagonalize the Hamiltonian via scattering matrix theory, and derive the conductance in the form of a Fano expression, which depends on the mean-field parameters. We predict that in the Kondo regime the magnetic field leads to a gapped energy level spectrum due to hybridisation of the non-interacting QD state and the Kondo state, and can quantum-mechanically alter the electron's path preference. We demonstrate that an abrupt symmetry change in the Fano resonance, as seen experimentally, could be a consequence of an underlying Kondo channel. 
\end{abstract}

\pacs{72.10.Fk, 73.21.La, 73.23.-b, 73.63.Kv}

\maketitle

\section{Introduction}\label{introduction}
One of the most fascinating manifestations of many-body physics in mesoscopic systems is the Kondo effect. It was originally observed in the 1930s in gold wires and formally explained in 1964 by Jun Kondo\cite{kondo_ptp32_37} in metals that contain magnetic impurities. The Kondo effect became an intense topic of research in the context of mesoscopic systems\cite{kouwenhoven_physw14_33} after its observation in quantum dots (QDs) in the late nineties.\cite{goldhaber_nature391_156,*cronenwett_science281_540} The Kondo effect in QDs is revealed as a zero-bias conductance enhancement at low temperatures when the QD is spin polarized and strongly coupled to the conduction electrons (such as a QD that is  tunnel-coupled to leads). Since its discovery in QDs, the Kondo effect has been observed and studied in a variety of exciting mesoscopic systems: individual atoms,\cite{otte} carbon nanotubes,\cite{nygard_nature408_342,*jarillo_nature434_484} molecules,\cite{
liang_nature417_725,*park_nature417_722} bucky 
balls\cite{yu_nanolett4_79} and graphene.\cite{chen_naturephys7_535}
\par Since the Kondo transport can strongly influence the Fano phenomenon, there has recently been a lot of interest in the interplay between these two effects in QDs. This interest comes from the use of Fano resonances as a tool to study coherence and to provide valuable transport information,\cite{verduijn_applphyslett96_072110,calvet_prb83_205415} that may result in future nano-scale device applications.\cite{cardamone_nanolett6_2422,darau_prb79_235404,busser_prb86_165410} The Fano effect results from the interference between a scattering event within a continuum of states (background process) and another one by means of a discrete energy level (resonant process). A Fano resonance manifests itself as an asymmetric conductance line-shape.\cite{fano} Fano resonances have been observed in many different physical systems and extensively studied in nanoscale systems.\cite{miroshnichenko_revmodphys82_2257}
Scientific awareness of the interaction between the Fano and the Kondo effect began with STM measurements on metal surfaces,\cite{li_prl80_2893} however, it found its greatest impact in mesoscopic systems. 
The Fano-Kondo effect is observed when Kondo transport takes over the background scattering process in the Fano effect. The trademark of such interaction is a conductance valley when tuning the energy level of the resonant process. 
\par So far, only a small number of dedicated Fano-Kondo studies have been reported.\cite{madhavan_science280_567,sato_prl95_066801,rushforth_prb73_081305,sasaki_physrevlett103_266806} \citet{sasaki_physrevlett103_266806} performed transport measurements through a side-coupled double QD configuration and found a strong modulation of the Kondo conductance by the Fano effect. Their results are in good agreement with a tight-binding model presented in a parallel theoretical publication.\cite{tamura_physe42_864} In a theoretical work, \citet{fang_prb81_113402} have calculated the conductance in an Aharonov-Bohm interferometer containing a Kondo QD coupled to a non-interacting QD. They predict a Fano-Kondo effect that could decoupled the two dots through the manipulation of the Kondo resonance by varying the inter-dot coupling. 
\par In this publication we investigate the Fano-Kondo interplay in an Aharonov-Bohm ring, containing a non-interacting QD and a Coulomb interacting QD. Using a scattering matrix approach combined with the slave-boson mean-field approximation (SBMF), we calculate the conductance and the Fano parameter. The derivation of the Fano form for the conductance allows us to analyze in-depth the Fano-Kondo phenomenon and, specially, the origin of the transport features. In addition, we present a qualitative comparison with the experiment performed by \citet{verduijn_applphyslett96_072110}, and suggest an explanation of their observations in terms of a Fano-Kondo interference phenomenon in an Aharonov-Bohm geometry. \par The paper is organized as follows: In section \ref{model} we introduce the Anderson Hamiltonian for a non-interacting QD and a Coulomb interacting QD connected in parallel. We diagonalize this Hamiltonian resulting in a set of self-consistent equations, which we solve numerically to obtain the 
conductance. In section \ref{results} we present our calculations. These lead us to predict a Fano footprint of an underlying Kondo channel and to show that the magnetic flux can open a gap between two resonances and alter the electron's path preference for transport.  We conclude in \ref{conclusion} with a discussion of applications of our results.

\section{Model $\&$ methodology}\label{model}
\begin{figure}[htbp]
\centering
\includegraphics[width=\columnwidth]{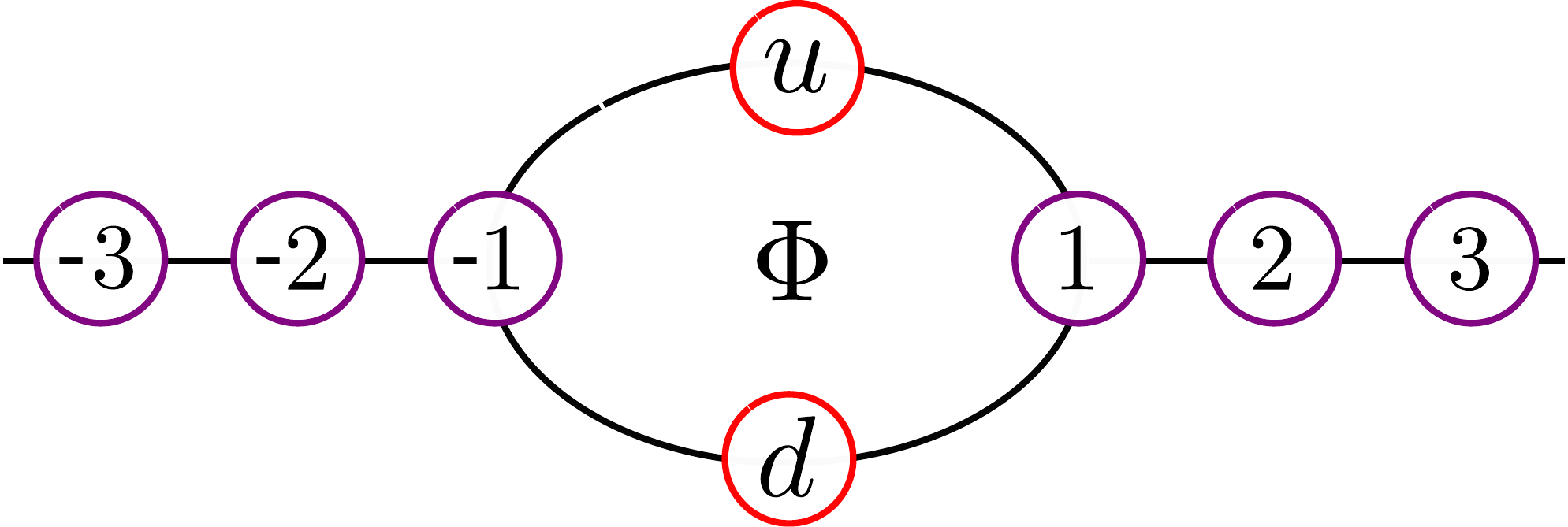}
\caption{Tight-binding model for two QDs embedded in an Aharonov-Bohm ring nanostructure. The QDs are labeled with indexes $u$ and $d$ for the upper and lower QD, respectively.}\label{ring} 
\end{figure}

We consider two QDs embedded in an Aharonov-Bohm ring which is attached to two 1D conducting leads. Fig. \ref{ring} shows a schematic representation of the nanostructure.
A magnetic field is used to create a magnetic flux ($\Phi$) through the ring, which gives a phase contribution ($\phi$) to the electron's wavefunction. The Hamiltonian of the system is given by
\begin{eqnarray}\label{hamiltonian}
H=&\epsilon_F&\sum_{j,\sigma}{c_{j,\sigma}^\dagger c_{j,\sigma}}\!-t_0\sum_{j,\sigma}(c_{j,\sigma}^\dagger c_{j+1,\sigma}+c.c.)\nonumber \\
&+&\sum_{n,\sigma}{\epsilon_n{f_{n,\sigma}^\dagger f_{n,\sigma}}}+U_df_{d\uparrow}^\dagger f_{d\uparrow}f_{d\downarrow}^\dagger f_{d\downarrow}\nonumber \\
&-&\sum_{n,\sigma}(t_{n,L}c_{-1,\sigma}^\dagger f_{n,\sigma}+t_{n,R}c_{1,\sigma}^\dagger f_{n,\sigma}+c.c).
\end{eqnarray}
We assume that each QD has one single energy level labeled $\epsilon_n$ (with $n=u$ for the upper QD and $n=d$ for the QD in the lower path). In the leads each site has a single-particle energy at the Fermi energy $\epsilon_F$. The hopping energy between neighboring sites in the leads is $t_0$. Taking into account the Aharonov-Bohm phase picked up by the electron ($\phi$), the hopping between the 1D leads and the ring structure is given by $t_{n,L}=t_ne^{\pm i\frac{\phi}{4}}$ and $t_{n,R}=t_ne^{\mp i\frac{\phi}{4}}$, where $t_{n,L}$ and $t_{n,R}$ ($n=u,d$), describe the hopping of an electron from the upper ($u$) or lower ($d$) QD to the left ($L$) or right ($R$) lead. The operators $c_{j,\sigma} (c_{j,\sigma}^\dagger)$ and $f_{n,\sigma} (f_{n,\sigma}^\dagger)$ correspond to the annihilation (creation) 
operators in the leads and in the QDs, respectively. Finally a Coulomb interaction is 
present 
in the lower QD with strength $U_d$. 
\par We calculate the conductance of this nanostructure in three regimes, depending on $\epsilon_d$, the energy level of the lower QD, 1) the Kondo regime ($\epsilon_d<\epsilon_F$, $U_d>\epsilon_F-\epsilon_d$ and $\Gamma_d<\epsilon_F-\epsilon_d$, where $\Gamma_d$ is the tunnel coupling of the lower QD to the leads), 2) the mixed-valence regime ($\epsilon_d<\epsilon_F$, $U_d>\epsilon_F-\epsilon_d$ and $\Gamma_d\approx\epsilon_F-\epsilon_d$) and 3) the empty orbital regime ($\epsilon_d>\epsilon_F$). For simplicity, we take $U_d\rightarrow \infty$ and use the infinite-U slave boson mean-field approximation to diagonalize the Hamiltonian $H$ [Eq. (\ref{hamiltonian})].\cite{coleman_physrevb29_3035,third} Within this scheme we introduce a boson creation (annihilation) operator $b^\dagger_d$($b_d$) that acts on the lower QD. Now the $b^\dagger_d\ket{0}$ state represents the empty state in the lower QD. We then have to create a boson when removing an electron from the QD to leave it in the 
empty state, $f_{d,\sigma}\rightarrow f_{d,\sigma}b_d^\dagger$. Since $U_d\rightarrow\infty$ double occupancy is forbidden which is imposed by the constraint $b_d^\dagger b_d+\sum_\sigma f_{d,\sigma}^\dagger f_{d,\sigma}=1$. Within this exact representation, the Hamiltonian $H$ of the system [Eq. (\ref{hamiltonian})] becomes:
\begin{eqnarray}\label{effective}
H_{SB}=&\epsilon_F&\sum_{j,\sigma}{c_{j,\sigma}^\dagger c_{j,\sigma}}\!-t_0\sum_{j,\sigma}(c_{j,\sigma}^\dagger c_{j+1,\sigma}+c.c.)\nonumber \\
&+&\sum_{n,\sigma}{\epsilon_n{f_{n,\sigma}^\dagger f_{n,\sigma}}}+\lambda_d(b_d^\dagger b_d + \sum_\sigma f_{d,\sigma}^\dagger f_{d,\sigma} -1)\nonumber \\
&-&\sum_\sigma(t_{u,L}c_{-1,\sigma}^\dagger f_{u,\sigma}+t_{u,R}c_{1,\sigma}^\dagger f_{u,\sigma}+c.c)\nonumber \\
&-&\sum_{\sigma}(t_{d,L}c_{-1,\sigma}^\dagger f_{d,\sigma}b_d^\dagger+t_{d,R}c_{1,\sigma}^\dagger f_{d,\sigma}b_d^\dagger+c.c).
\end{eqnarray}
Here the constraint has been added by means of a Lagrange multiplier, $\lambda_d$. Since the Hamiltonian $H_{SB}$ [Eq.  (\ref{effective})] remains difficult to solve, we use the mean-field approximation were fluctuations around the average value of the boson operators in Eq.  (\ref{effective}) are neglected. By substituting the boson operators with their real expectation values, $b_d\rightarrow<b_d>$, the slave-boson Hamiltonian [Eq. (\ref{effective})] is replaced by an effective mean-field Hamiltonian,
\begin{eqnarray}\label{effective_meanfield}	
H_{SB,MF}=&\epsilon_F&\sum_{j,\sigma}{c_{j,\sigma}^\dagger c_{j,\sigma}}\!-t_0\sum_{j,\sigma}(c_{j,\sigma}^\dagger c_{j+1,\sigma}+c.c.)\nonumber \\
&+&2(\epsilon_d+\lambda_d)f_d^\dagger f_d+\lambda_d(<b_d>^2-1)\nonumber \\
	&+&2\epsilon_uf_u^\dagger f_u-2(t_{u,L}c_{-1}^\dagger f_u+t_{u,R}c_1^\dagger f_u+c.c)\nonumber \\
	&-&2<b_d>(t_{d,L}c_{-1}^\dagger f_d+t_{d,R}c_1^\dagger f_d+c.c).
\end{eqnarray}
The factor of 2 takes into account the contribution from both spin directions.  We now minimize the ground state energy of Eq. (\ref{effective_meanfield}) with respect to the Lagrange multiplier, $\lambda_d$, and the boson expectation value, $<b_d>$,
\begin{subequations}\label{minimization}
\begin{eqnarray}
\pd{<H_{SB,MF}>}{\lambda_d}&=&2<f^\dagger_df_d>+<b_d>^2-1=0, \\
\pd{<H_{SB,MF}>}{<b_d>}&=&\lambda_d <b_d>^2-R_d=0,
\end{eqnarray}
\end{subequations}
where
\begin{eqnarray}
R_d=&<&b_d>(t_{d,L}<c^\dagger_{-1}f_d>+t_{d,R}<c^\dagger_1f_d>\nonumber \\
&+&t_{d,L}^*<f^\dagger_dc_{-1}>+t_{d,R}^*<f^\dagger_dc_1>).
\end{eqnarray}
The fermion expectation values in Eq. (\ref{minimization}) are obtained by diagonalizing the mean-field Hamiltonian [Eq. (\ref{effective_meanfield})]. We take as ansatz the linear combination of atomic orbitals, $\ket{\Psi_k}=\sum_n a_{n}^k\ket{n}+\sum_j a_{j}^k\ket{j}$, where $a_{j}^k$ and $a_{n}^k$ are the probability amplitudes to find the electron, with momentum $k$ and energy $w=\epsilon_F-2t_0\cos{(k)}$, at site $j$ in the leads and at quantum dot $n=u,d$, respectively.
We assume that the transport of electrons is described by an incoming plane wave that is reflected and transmitted at the Aharonov-Bohm nanostructure. Assuming an incoming electron from the left lead ($L$), where the chemical potential is at $\epsilon_F+V_L$, yields: $a_{jL}^k=e^{ik\cdot j}+r_{LL}e^{-ik\cdot j}$ for $j<0$ and $a_{jL}^k=\tau_{RL}e^{ik\cdot j}$ for $j>0$, where $0\leq k\leq k_L=\arccos(\frac{V_L}{2t_0})$. $\tau_{RL}$ and $r_{LL}$ correspond to the transmission amplitude from the left to the right lead and the reflection amplitude in the left lead, respectively.\cite{orellana_physrevb74_193315,sadreev_jphysmat36_11413}. 
\par By substituting the ansatz, $\ket{\Psi_k}$, into the Schr\"{o}dinger equation using the mean-field Hamiltonian [Eq. (\ref{effective_meanfield})], we arrive at the following system of linear equations:
\begin{subequations}\label{system}
\begin{eqnarray}
-t_0r_{LL}+t_ue^{i\frac{\phi}{4}}a_{uL}+\tilde{t}_de^{-i\frac{\phi}{4}}a_{dL}&=&t_0\\	
-t_0\tau_{RL}+t_ue^{-i\frac{\phi}{4}}a_{uL}+\tilde{t}_de^{i\frac{\phi}{4}}a_{dL}&=&0\\		
t_ue^{-i\frac{\phi}{4}}r_{LL}+t_ue^{i\frac{\phi}{4}}\tau_{RL}\hspace{14mm}&&\nonumber \\
+(w-\epsilon_u)e^{-ik}a_{uL}&=&-t_ue^{-i\frac{\phi}{4}}e^{-2ik}\\		
\tilde{t}_de^{i\frac{\phi}{4}}r_{LL}+\tilde{t}_de^{-i\frac{\phi}{4}}\tau_{RL}\hspace{14mm}&&\nonumber \\
+(w-\tilde{\epsilon}_d)e^{-ik}a_{dL}&=&-\tilde{t}_de^{i\frac{\phi}{4}}e^{-2ik},
\end{eqnarray}
\end{subequations}
where we have renormalized the coupling and the energy level of the lower QD as $\tilde{t}_d=<b_d>t_d$ and $\tilde{\epsilon}_d=\epsilon_d+\lambda_d$, respectively. By solving Eq. (\ref{system}) we obtain all the probability amplitudes, $a_{jL}^k$ and $a_{nL}^k$, for an incoming wave from the left. In the same way we solve for an incoming wave from the right lead, where the chemical potential is at $\epsilon_F+V_R$, by using the coefficients $a_{jR}^k=e^{-ik\cdot j}+r_{RR}e^{ik\cdot j}$ for $j>0$ and $a_{jR}^k=\tau_{LR}e^{-ik\cdot j}$ for $j<0$, where $0\leq k\leq k_R=\arccos(\frac{V_R}{2t_0})$. Since the magnetic field breaks the time reversal symmetry, coefficients and probability amplitudes for a left and right incoming wave are different. With all the coefficients we can compute the fermion expectation values in Eq. (\ref{minimization}).\cite{orellana_physrevb74_193315} We obtain the boson parameter $<b_d>$ and Lagrange multiplier $\lambda_d$ from the self-consistent set of equations [Eqs. (\ref{
minimization}
-\ref{system})], at equilibrium and 
zero temperature. From the transmission coefficient in Eq. (\ref{system}) the conductance at the Fermi energy ($k=\hspace{.1mm}^\pi\!\!/\!_2$) reads $G=\frac{2e^2}{h}\abs{\tau_{RL}({^\pi\!\!/\!_2})}^2$, with
\begin{equation} \label{transmission}
\tau_{RL}({^\pi\!\!/\!_2})=-\frac{\epsilon_u\tilde{\Gamma}_de^{i\frac{\phi}{2}}+\tilde{\epsilon}_d\Gamma_ue^{-i\frac{\phi}{2}}}{[\epsilon_u-i\Gamma_u][\tilde{\epsilon}_d-i\tilde{\Gamma}_d]+\Gamma_u\tilde{\Gamma}_d\cos^2(\frac{\phi}{2})}.
\end{equation}
In Eq. (\ref{transmission}) we have set the Fermi level to zero ($\epsilon_F=0$) and introduced the coupling strengths $\Gamma_u=2\frac{t_u^2}{t_0}$ and $\tilde{\Gamma}_d=2\frac{\tilde{t}_d^2}{t_0}$. Eq. (\ref{transmission}) reduces to the transmission calculated by \citet{lopez_prb71_115312} for identical QDs and symmetric junctions. 
\par Having obtained the ingredients to calculate the conductance in the presence of Coulomb interaction in the lower QD, we now consider the regime in which the lower arm of the Aharonov-Bohm ring is tuned to the Kondo regime and scattering occurs within a continuum (background process). In the upper arm transport occurs through the discrete QD energy level (resonant process). The interference between both processes will give rise to a Fano line-shape in the conductance. The transport is then characterized by the Fano parameter $q$, which in general is a complex number. The parameter $\abs{q}$ measures the ratio of the resonant scattering to the background scattering amplitude. 
When $\abs{q}>>1$ the electronic transport is dominated by the resonant process and the conductance takes a Lorentzian shape. As the background process starts to contribute, the Fano parameter increases. When both processes contribute significantly $\abs{q}\approx 1$, and the line-shape becomes asymmetric. When the background process takes over as the main mechanism for transport $\abs{q}\approx 0$ and the conductance shows a symmetrical dip centered at the resonance.\cite{miroshnichenko_revmodphys82_2257} Below, we show that by tuning the energy level of the interacting QD we can set the conductance to any of the line-shapes mentioned before, and that interference tuned by the magnetic flux leads to interesting behavior of the transmission. First we use Eq. (\ref{transmission}) to write the conductance, in terms of $\epsilon_u$, in a Fano form:\cite{fano}
\begin{equation}\label{fanoform}
G=G_o\frac{\abs{(\epsilon_u+\epsilon_o)+q\Gamma}^2}{(\epsilon_u+\epsilon_o)^2+\Gamma^2},
\end{equation}
where the amplitude $G_o$, the resonant point $\epsilon_o$, the peak broadening $\Gamma$ and the Fano parameter $q$ are given by:
\begin{subequations}\label{fanoexpression}
\begin{eqnarray}
	G_o&=&\frac{2e^2}{h}\frac{1}{\eta^2+1},\\
	\epsilon_o&=&\Gamma_u\frac{\eta}{\eta^2+1}\cos^2(\frac{\phi}{2}),\\	\Gamma&=&\Gamma_u\frac{\eta^2+\sin^2(\frac{\phi}{2})}{\eta^2+1}\label{gamma},\\			 			   		q&=&\frac{\eta[\eta^2\cos(\phi)-\sin^2(\frac{\phi}{2})-i(\eta^2+1)\sin(\phi)]}{\eta^2+\sin^2(\frac{\phi}{2})}.\label{fanoparameter}
\end{eqnarray}
\end{subequations}
The parameter $\eta=\frac{\tilde{\epsilon}_d}{\tilde{\Gamma}_d}$ is a measure of the Kondo effect strength in the lower QD. Eq. (\ref{fanoform}) reveals a shift by $\epsilon_o$ of the center of the resonance away from its bare value $\epsilon_u=\epsilon_F=0$, as a consequence of interference.

\section{Calculations $\&$ results}\label{results}

In this section we analyze the result for the conductance [Eq. (\ref{fanoform})] from the previous section by solving the system of equations [Eqs. (\ref{minimization})-(\ref{system})] numerically. Unless stated otherwise we take the wide band limit and the symmetric case: $\Gamma_u=\Gamma_d=\frac{t_o}{100}$, where $t_o$ is taken as the unit of energy. 
\subsection{Kondo channel as continuum scattering path}
\begin{figure}[htbp]
\centering
\includegraphics[width=\columnwidth]{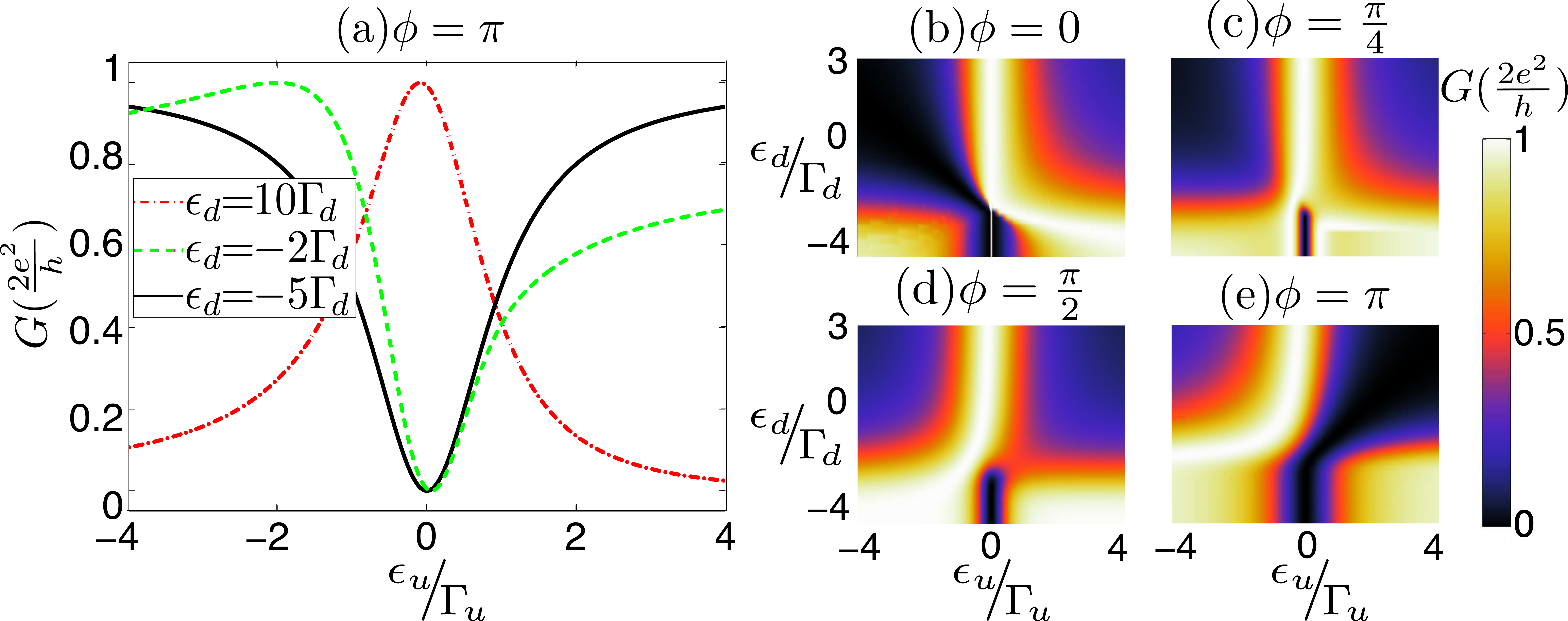}
\caption{(a) Normalized conductance, $G$, for the three different regimes: Lorentzian (resonant channel dominates, red dot-dashed curve), mixed-valence (resonant channel and Kondo channel contribute, purple dashed curve) and Fano-Kondo (Kondo channel dominates, black continuous curve). (b)-(e) Normalized conductance, $G$, versus energy levels, $\epsilon_u$ and $\epsilon_d$, for different magnetic fluxes through the ring.}\label{regimes} 
\end{figure}

We begin by showing that our calculations indicate that the Kondo channel through the lower QD is acting as the continuum scattering path within the Fano effect. Fig. \ref{regimes}(a) shows the different line-shapes for the three regimes mentioned in section \ref{model}. When $\epsilon_d>>\Gamma_d$ the probability of an electron at the Fermi level to tunnel to the lower QD is negligible. The conductance is therefore determined by electrons transversing the upper QD. 
This gives a resonance with a Lorentzian shape centered at the Fermi level $\epsilon_u=0$ [red point-dashed curve in Fig. \ref{regimes}(a)] for an arbitrary magnetic phase shift $\phi$ [Figs. \ref{regimes}(b)-\ref{regimes}(e)]. In this regime there is no Kondo effect, $\eta>>1$, and Eq. (\ref{fanoform}) reduces to:
\begin{equation}\label{resonanteq}
 G=\Big(\frac{2e^2}{h}\Big)\frac{1}{\eta^2} \frac{\abs{\epsilon_u+\eta e^{-i\phi}\Gamma_u}^2}{\epsilon_u^2+\Gamma_u^2}.
\end{equation}
We also find $\abs{q}>>1$, hence the system behaves as if there were no lower path in the ring and transport is dominated by the resonant channel. 
\par We now consider the interacting QD to be in the mixed-valence regime $\epsilon_d\approx-\Gamma_d$. Here, both arms contribute to the transmission through the structure. The interference between the resonant upper path and the mixed-valence lower path, acting as a continuum, gives rise to a Fano shape [dashed curve in Fig. \ref{regimes}(a)] with its symmetry tuned by the magnetic phase shift $\phi$ [Figs. \ref{regimes}(b)-\ref{regimes}(e)]. This symmetry is obtain by computing the argument of the Fano parameter\cite{verduijn_applphyslett96_072110} [Eq. (\ref{fanoparameter})]:
\begin{equation}\label{argument}
  \arg(q)=\arctan{\Bigg[\frac{2}{\tan{(\frac{\phi}{2})}-\frac{\eta^2}{\eta^2+1}\cot{(\frac{\phi}{2})}}\Bigg]}.
\end{equation}
In this regime the Fano parameter gives $\abs{q}\approx1$, which agrees with the interpretation of $\abs{q}$ [for the green dashed curve in Fig. \ref{regimes}(a) $\abs{q}=1.2$]. 
\subsection{Symmetry change due to Fano-Kondo interplay}
\begin{figure}[htbp]
\centering
\includegraphics[width=\columnwidth]{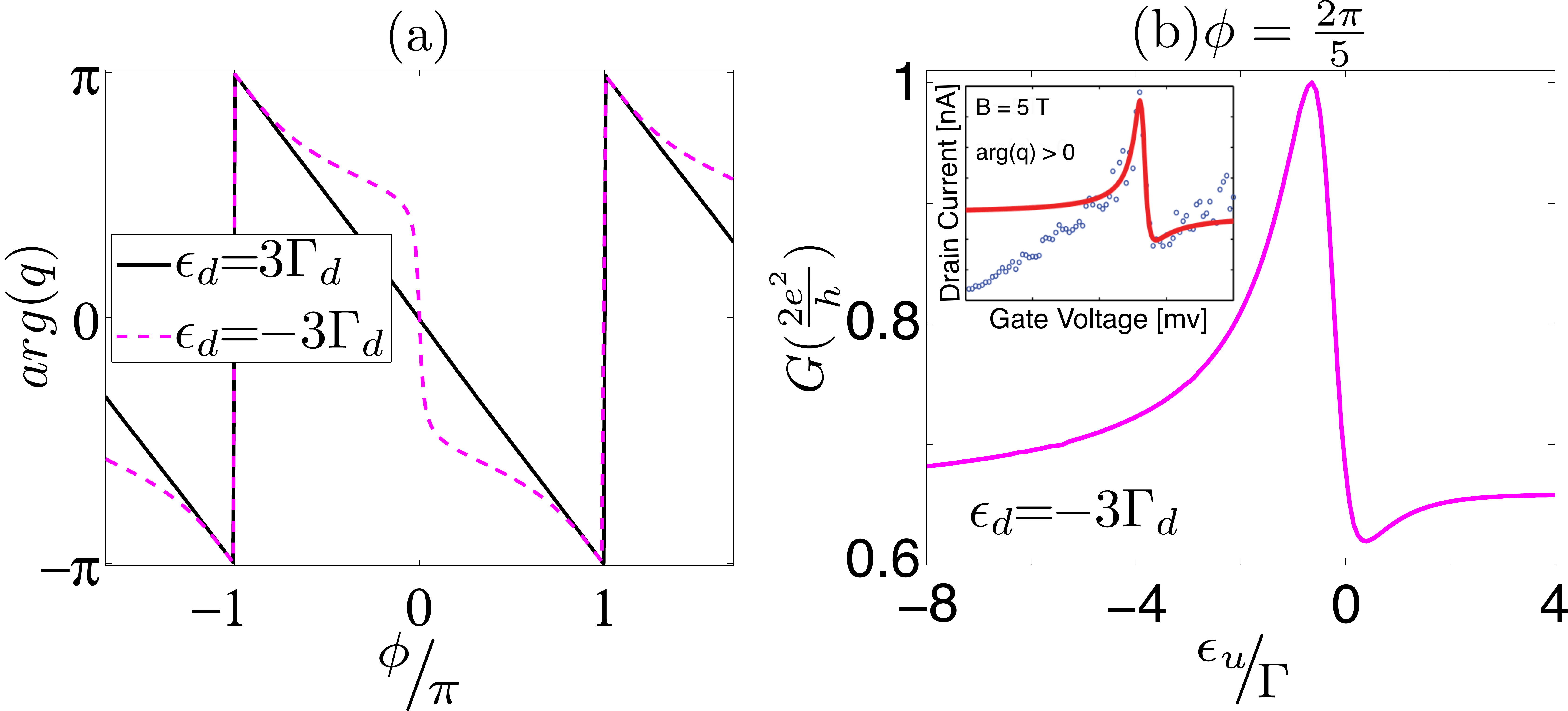}
\caption{Calculations performed for $\Gamma_d=10\Gamma_u$. (a) Degree of symmetry of the resonance obtained from Eq. (\ref{argument}) for $\epsilon_d=3\Gamma_d$ (black continuous) and $\epsilon_d=-3\Gamma_d$ (green dashed). (b) Calculated conductance for an energy level of the interacting QD as in the experiment ($\epsilon_d=-3\Gamma_d$). Inset in (b) shows the reported data and fit of the experiment [J. Verduijn \emph{et al.}, Appl. Phys. Lett. \textbf{96},
072110 (2010)].}\label{mixed} 
\end{figure}

In a transport experiment performed by \citet{verduijn_applphyslett96_072110} using a three-dimensional silicon field effect transistor (FinFET), where a few arsenic dopants diffuse to the transistor channel, an asymmetric Fano resonance for one of the arsenic atoms was observed at zero-bias voltage, and speculated that the Fano resonance results from the interference between a Kondo transport channel and a direct transport process.  The geometry consists of two parallel paths with an impurity in each one. Using parameters similar to the ones in this experiment, $\Gamma_d=10\Gamma_u$ and $\epsilon_d=-3\Gamma_d$, we obtain qualitative agreement. Fig. \ref{mixed}(b) shows the calculated conductance 
for a magnetic phase shift of $2\pi/5$, in order to obtain a comparable degree of symmetry with the reported resonance [inset in Fig. \ref{mixed}(b)]. In this calculation, a value of $\abs{q}=1.14$ was obtained, which indicates that the measured asymmetric 
conductance profile originates from an underlying mixed-valence Kondo path to the upper QD bare resonance. The degree of symmetry [$\arg(q)$] versus magnetic field was calculated in the experiment, and an abrupt jump in symmetry at half a period was found.\cite{verduijn_applphyslett96_072110} We recognize this sudden symmetry change as a trademark of an underlying Kondo channel of the transport process. Fig. \ref{mixed}(a) shows the argument of the calculated Fano parameter [Eq. (\ref{argument})] for $\Gamma_d=10\Gamma_u$. A linear change in symmetry occurs when the lower QD is completely outside the Kondo regime [black continuous line in Fig. \ref{mixed}(a)]. As soon as the QD enters the Kondo regime (even in the mixed-valence regime) an abrupt symmetry change begins to form at half a period [green dashed line in Fig. \ref{mixed}(a)]. This transition becomes more abrupt as the lower QD enters deeper into the Kondo regime ($\eta<<1$).\cite{first}
\par In the Fano-Kondo regime the lower QD is tuned to the Kondo state by lowering its energy level to $\epsilon_d<<-\Gamma_d$ [black continuous curve in Fig. \ref{regimes}(a)]. The shape of the conductance then changes from a peak to a dip line-shape. As was observed experimentally,\cite{li_prl80_2893} this conductance behavior indicates that the Kondo channel has taken over the role of scattering within a continuum in a Fano effect, interfering with a discrete scattering process mediated by the upper QD. For a magnetic flux away from $\phi=2\pi n$, transport is dominated by the Kondo channel regardless of the degree of interference [Figs. \ref{regimes}(b)-\ref{regimes}(e)].\cite{miroshnichenko_revmodphys82_2257,second} Within this regime the lower QD is in the Kondo state, then $\lambda\rightarrow-\epsilon_d$ such that $\eta<<1$. In this limit $|q|=\eta<<1$ and
\begin{equation}\label{fano-kondo}
 G=\Big(\frac{2e^2}{h}\Big)\frac{\abs{\epsilon_u}^2}{\epsilon_u^2+[\Gamma_u\sin^2{(\frac{\phi}{2})}]^2}.
\end{equation}
In contrast with the constant broadening of the peak shape [red point-dashed curve in Fig. \ref{regimes}
(a)] the broadening of the dip is modulated by $\sin^2{(\frac{\phi}{2})}$ as seen in Figs. \ref{regimes}(b)-\ref{regimes}(e). Furthermore, the expression for $q$ [Eq. (\ref{fanoparameter})] is in accordance with the appropriate interpretation of the phenomenon [for the black continuous curve in Fig. \ref{regimes}(a) $\abs{q}=0.3$]. 
\subsection{Tuning the electron's path preference}
So far, we have seen that our system can be in three different regimes depending on the relative position of the energy level of the interacting QD. Our results show that the magnetic flux can be used as a tool to tune the conductance within each of the regimes with characteristic landmarks: no effect in the resonant regime, symmetry change of Fano resonance in the mixed-valence regime and change of the dip broadening in the Fano-Kondo regime. It is in this last regime where the magnetic flux can have the extra role of tuning the gap between two resonances and influences not only the interference at the end of the ring, but affects the probability of the electron taking the lower or upper path.
\begin{figure}[htbp]
\centering
\includegraphics[width=\columnwidth]{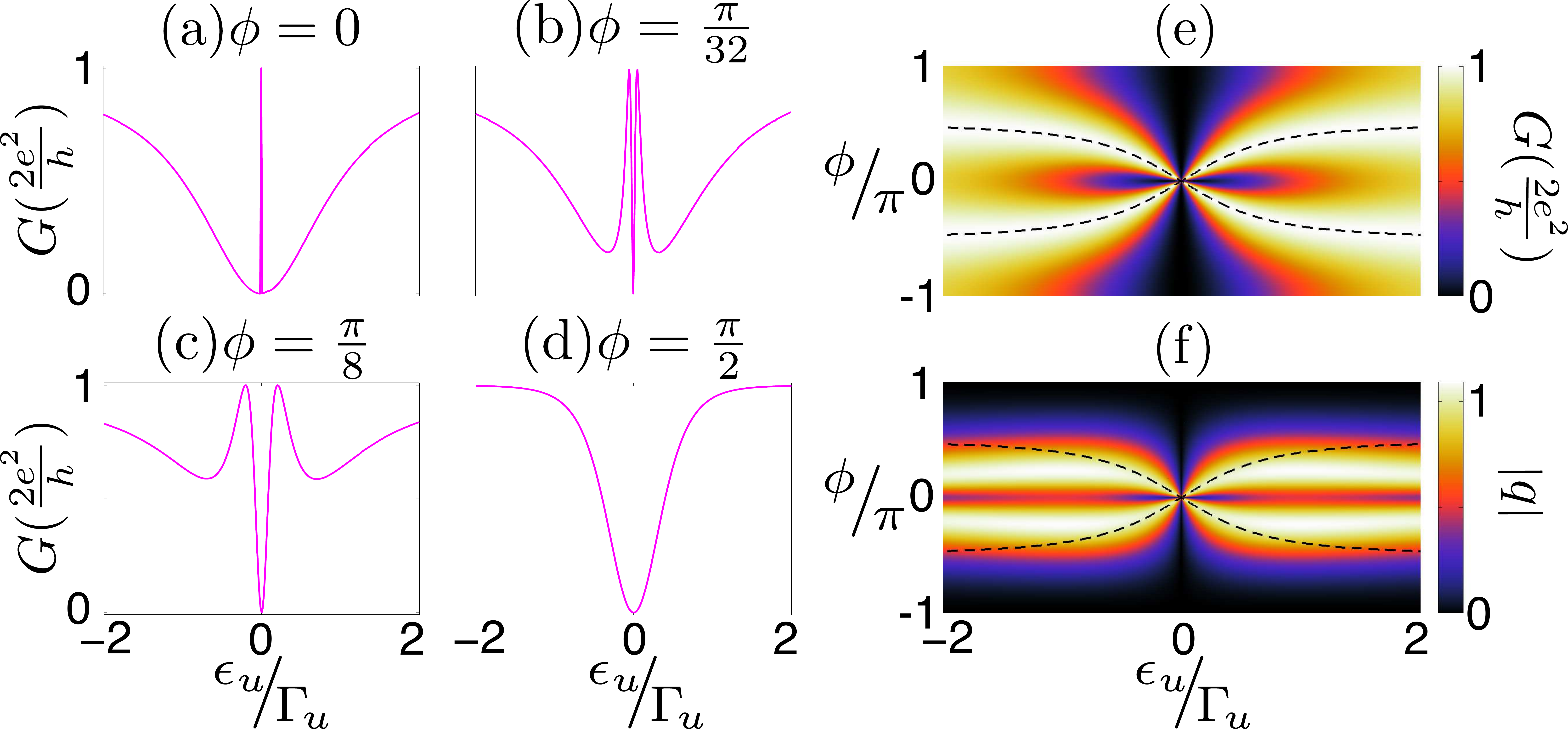}
\caption{Calculations perfomed for $\epsilon_d=-5\Gamma_d$. (a-d) Conductance, $G$, versus the noninteracting QD energy level, $\epsilon_u$, for different magnetic phase shifts, $\phi$, close to zero. (e) Conductance, $G$, versus $\epsilon_u$ and magnetic phase $\phi$. Figs. (a)-(d) are different horizontal cuts of (e). (f) Absolute value of the Fano parameter [Eq. (\ref{fanoparameter})] for the same parameters as in (e).}\label{kondo} 
\end{figure}
\par When $\phi\approx 2n\pi$ the conductance [Eq. (\ref{fanoform})] cannot be written as Eq. (\ref{fano-kondo}). Even though the conductance for small magnetic shifts $\phi$ [Figs. \ref{kondo}(a)-\ref{kondo}(c)] has a distinctive line-shape, the antiresonance near $\epsilon_u=0$, of width $\Gamma\propto\tilde{\epsilon}_d^2+\tilde{\Gamma}_d^2\sin^2{(\frac{\phi}{2})}$, remains present for any value of magnetic flux. This magnetic flux independence is due to the fact that near $\epsilon_u=0$ the upper discrete path is close to resonance, such that the transport characteristics through it remain robust under the influence of a magnetic flux. Therefore, the only effect remaining is the interference with the Kondo channel, giving the already mentioned dip line-shape [Figs. \ref{kondo}(a)-\ref{kondo}(c)]. Moreover, two sharp resonances appear when $\phi\approx 2n\pi$ [Figs. \ref{kondo}(a)-\ref{kondo}(c)]. Since these two resonances originate from the contribution of the discrete energy level path and the Kondo 
path [$\abs{q}\approx 1$ in Fig. \ref{kondo}(f) as indicated by the contour dashed line mapping both resonances], we speculate that they appear because of a hybridization of the non-interacting QD state with the Kondo state, which leads to a gap opening between two hybrid states. This gapped system could lead to interesting physics as we control the state of the system (under or above the gap) as well as the size of the gap, by tuning the non-interacting QD energy level and the magnetic flux, respectively.

\par We can also use the magnetic field to alter the electron's path preference, even if the interacting QD is deep in the Kondo regime and this path becomes the preferred channel for the scattering process. Fig. \ref{kondo}(e) shows how the conductance behavior changes near $\phi=0$ and the corresponding absolute value of the Fano parameter [Fig. \ref{kondo}(f)] indicates that the transport mechanism changes from dominance of the Kondo channel ($\abs{q}\approx0$) to a mechanism in which both scattering paths contribute ($\abs{q}\approx 1$). The effect is a consequence of the decrease of the Kondo temperature via the manipulation of the Kondo-QD tunnel coupling by the magnetic flux. The tunnel coupling decreases until it becomes equally probable to tunnel to either QD, such that both scattering paths contribute significantly to the conductance. The dependence of the Kondo temperature on magnetic flux can be derived by writing the Fano form for conductance [Eq. (\ref{fanoform})] in terms of the interacting 
energy level $\tilde{\epsilon}_d$, 
such that the expression found will describe the Kondo 
resonance. Then, the derived state broadening ($\Gamma$) is substituted into the expression for the Kondo temperature using SBMF Hamiltonian,\cite{oguchi} 
\begin{equation}
k_BT_K=\sqrt{\tilde{\epsilon}_d^2+\tilde{\Gamma}_d^2\left[\frac{\zeta^2}{\zeta^2+1}+\frac{\sin^2\left(\frac{\phi}{2}\right)}{\zeta^2+1}\right]^2},
\end{equation}
with $\zeta=\frac{\epsilon_u}{\Gamma_u}$, $k_B$ the Boltzman constant and $T_K$ the Kondo temperature.

\section{Conclusions}\label{conclusion}
In this paper, we have studied coherent transport through an Aharonov-Bohm ring structure with two embedded QDs. In particular, we have analyzed the influence of a Coulomb interacting channel to a discrete scattering transport process in an Aharonov-Bohm configuration. Three regimes were identified, depending on the contribution of the Kondo channel to the conductance. These regimes yield different conductance line-shapes: Lorentzian, asymmetric Fano and symmetric dip. The influence of a magnetic phase shift to these regimes is to tune the Fano resonance asymmetry and to tune the broadening of the symmetrical dip. We have shown that the Fano resonance line-shape can be explained by an underlying scattering mechanism through a Kondo channel, and that the shape can vary from an asymmetric resonance to a symmetric dip as we drive the Coulomb blocked QD further into the Kondo regime. We have also shown that we can identify the presence of an underlying Kondo channel by 
an abrupt change in symmetry of the discrete channel resonance. Furthermore, we have found agreement with experimental data that support our findings. Finally, we predict a gap opening between two resonances that we speculate is a consequence of gapped hybrid states. We also predict that, even if the interacting channel is fully in the Kondo regime, we can use the magnetic flux to diminish its contribution by lowering the characteristic Kondo temperature (Kondo state broadening), producing an alteration in the electron's path preference.

\section*{Acknowledgments}
This work is part of the research program of the Foundation for Fundamental Research on Matter (FOM), which is part of the Netherlands Organization for Scientific Research (NWO). This research was financially supported by the ARC Centre of Excellence for Quantum Computation and Communication Technology (CE110001027) and the Future Fellowship (FT100100589).
\bibliography{biblio}

\begin{thebibliography}{33}
\bibitem [{thi()}]{third}%
  \BibitemOpen
  \href@noop {} {}\bibinfo {note} {Even though we have limited our theoretical
  treatment to an infinite-U approach, the conclusions made also hold for
  finite-U because of the equivalence between both approaches: a strong Kondo
  channel in the presented framework, $\epsilon_d<<\epsilon_F$, is equivalent
  to having a finite-U system near the particle-hole symmetry point,
  $\epsilon_F-\epsilon_d=\frac{U}{2}$.}\BibitemShut {Stop}%
\bibitem [{fir()}]{first}%
  \BibitemOpen
  \href@noop {} {}\bibinfo {note} {This can be seen from Eq.(\ref{argument}).
  Outside the Kondo regime $\eta>>1$ then $\frac{\eta^2}{\eta^2+1}\approx1$,
  this means that $\arg(q)=\pm\frac{\pi}{2}$ at $\phi=\mp m\frac{\pi}{2}$ with
  $m$ odd. As we take the QD to the Kondo regime $\frac{\eta^2}{\eta^2+1}<1$
  and $\cot(\frac{\phi}{2})$ needs to compensate to obtain
  $\arg(q)=\pm\frac{\pi}{2}$. Then as $\eta\rightarrow0$,
  $\lim\limits_{\phi\rightarrow2z\pi^\mp}\arg(q)=\pm\frac{\pi}{2}$ with $z$
  integer, creating an abrupt symmetry change near $\phi=2\pi z$.}\BibitemShut
  {Stop}%
\bibitem [{sec()}]{second}%
  \BibitemOpen
  \href@noop {} {}\bibinfo {note} {When the magnetic phase is
  $\phi\approx2n\pi$, with $n$ an integer, there is a battle between the limit
  $\tilde{\epsilon}_d\rightarrow0$ which gives $\eta<<1$ and the limit
  $\phi\rightarrow2n\pi$ which gives $\sin(\frac{\phi}{2})<<1$. In this case
  the conductance cannot be reduce to a simple Lorentzian expression
  [Eq.(\ref{fano-kondo})].}\BibitemShut {Stop}%
\end{thebibliography}
\bibliographystyle{apsrev4-1}

\end{document}